\documentclass[pra,aps,showpacs,superscriptaddress]{revtex4}
\usepackage[T1]{fontenc}
\usepackage{textcomp}
\usepackage{bm}
\usepackage{amsmath}
\usepackage{graphicx}
\usepackage{esint}

\begin{document}

\title{Quantum metrology with unitary parametrization processes}

\author{Jing Liu}
\affiliation{Zhejiang Institute of Modern Physics, Department of Physics, Zhejiang
University, Hangzhou 310027, China}

\author{Xiao-Xing Jing}
\affiliation{Zhejiang Institute of Modern Physics, Department of Physics, Zhejiang
University, Hangzhou 310027, China}

\author{Xiaoguang Wang}
\email{xgwang@zimp.zju.edu.cn}

\affiliation{Zhejiang Institute of Modern Physics, Department of Physics, Zhejiang
University, Hangzhou 310027, China}
\affiliation{Synergetic Innovation Center of Quantum Information and Quantum Physics,
University of Science and Technology of China, Hefei, Anhui 230026, China}

\begin{abstract}
Quantum Fisher information is a central quantity in quantum metrology.
We discuss an alternative representation of quantum Fisher information
for unitary parametrization processes. In this representation,
all information of parametrization transformation, i.e., the
entire dynamical information, is totally involved in a Hermitian operator
$\mathcal{H}$. Utilizing this representation, quantum Fisher information
is only determined by $\mathcal{H}$ and the initial state.
Furthermore, $\mathcal{H}$ can be expressed in an expanded form. The highlights
of this form is that it can bring great convenience during the calculation for the Hamiltonians owning recursive commutations with their partial derivative.
We apply this representation in a collective spin system and show the specific
expression of $\mathcal{H}$. For a simple case, a spin-half system,
the quantum Fisher information is given and the optimal states to
access maximum quantum Fisher information are found. Moreover,
for an exponential form initial state, an analytical expression of
quantum Fisher information by $\mathcal{H}$ operator is provided.
The multiparameter quantum metrology is also considered and discussed
utilizing this representation.
\end{abstract}

\maketitle

How to precisely measure the values of physical quantities, such as
the phases of light in interferometers, magnetic strength, gravity
and so on, is always an important topic in physics. Obtaining high-precision
values of these quantities will not only bring an obvious advantage
in applied sciences, including the atomic clocks, physical geography,
civil navigation and even military industry, but also accelerate the
development of fundamental theories. One vivid example is the search
for gravitational waves. Quantum metrology is such a field attempting
to find optimal methods to offer highest precision of a parameter
that under estimation. In recently decades, many protocols and strategies
have been proposed and realized to improve the precisions of various
parameters~\cite{1,2,3,4,5,6,7,8,9,10,11,12,13,14,Yao,Xiao,
15,16,Escher,Vidrighin,Braun,Giovannetti06,Boixo,Taddei}.
Some of them can even approach to the Heisenberg limit, a limit given
by the quantum mechanics, showing the power of quantum metrology.

Quantum Fisher information is important in quantum metrology because
it depicts the theoretical lowest bound of the parameter's variance
according to Cram\'{e}r-Rao inequality~\cite{Helstrom,Holevo}. The
quantum Fisher information for parameter $\alpha$ is defined as $F=\mathrm{Tr}(\rho L^{2})$,
where $\rho$ is a density matrix dependent on $\alpha$ and $L$
is the symmetric logarithmic derivative (SLD) operator and determined
by the equation $\partial_{\alpha}\rho=(\rho L+L\rho)/2$. For a multiparameter
system, the counterpart of quantum Fisher information is called quantum
Fisher information matrix $\mathcal{F}$, of which the element is defined as $\mathcal{F}_{\alpha\beta}=\mathrm{Tr}(\rho\{L_{\alpha},L_{\beta}\})$,
where $L_{\alpha}$, $L_{\beta}$ are the SLD operators for parameters
$\alpha$ and $\beta$, respectively.

Recently, it has been found~\cite{Pang} that quantum Fisher information
can be expressed in an alternative representation, that all information
of parametrization process in quantum Fisher information is involved
in a Hermitian operator $\mathcal{H}$. This operator characterizes
the dynamical property of the parametrization process, and totally
independent of the selection of initial states. Utilizing this representation,
the quantum Fisher information is only determined by $\mathcal{H}$
and the initial state.

In this report, we give a general expression of quantum Fisher information
and quantum Fisher information matrix utilizing $\mathcal{H}$ operator.
For a unitary parametrization process, $\mathcal{H}$ can be expressed
in an expanded form. This form is particularly useful when the Hamiltonian owns
a recursive commutation relation with its derivative on parameter estimation.
We calculate the specific expression of $\mathcal{H}$
in a collective spin system, and provide an analytical expression
of quantum Fisher information in a spin-half system for any initial
state. Based on this expression, all optimal states to access maximum
quantum Fisher information are found in this system. Furthermore,
considering this spin-half system as a multiparameter system, the
quantum Fisher information matrix, can be easily obtained by the known form of $\mathcal{H}$ in single parameter estimations. On the other hand, inspired by a recent work~\cite{Jiang14},
for an exponential form initial state, we provide an analytical expression
of quantum Fisher information using $\mathcal{H}$ operator. A demonstration
with a spin thermal initial state is given in this scenario. The maximum
quantum Fisher information and the optimal condition are also discussed.

\vspace{8pt}
\noindent\textbf{Results}\\
\textbf{Quantum Fisher information with $\mathcal{H}$ operator.} For a general unitary parametrization
transformation, the parametrized state $\rho(\alpha)$ is expressed
by $\rho(\alpha)=U(\alpha)\rho_{0}U^{\dagger}(\alpha)$, where $\rho_{0}$
is a state independent of $\alpha$. In this paper, since the parameter
$\alpha$ is only brought by $U(\alpha)$, not the initial state $\rho_{0}$,
we use $U$ instead of $U(\alpha)$ for short. Denote the spectral decomposition of $\rho_{0}$
as $\rho_{0}=\sum_{i=1}^{M}p_{i}|\psi_{i}\rangle\langle\psi_{i}|$,
where $p_{i}$ and $|\psi_{i}\rangle$ are the $i$th eigenvalue and
eigenstate of $\rho_{0}$ and $M$ is the dimension of the support
of $\rho_{0}$. It is easy to see that $p_{i}$ and $U|\psi_{i}\rangle$
are the corresponding eigenvalue and eigenstate of $\rho(\alpha)$,
respectively. The quantum Fisher information for $\rho(\alpha)$ can
then be expressed by~\cite{Commuliu,PhysicaAliu}
\begin{equation}
F=\sum_{i=1}^{M}4p_{i}\langle\Delta^{2}\mathcal{H}\rangle_{i}-\sum_{i\neq j}\frac{8p_{i}p_{j}}{p_{i}+p_{j}}|\langle\psi_{i}|\mathcal{H}|\psi_{j}\rangle|^{2},\label{eq:F}
\end{equation}
where
\begin{equation}
\mathcal{H}:=i(\partial_{\alpha}U^{\dagger})U
\end{equation}
is a Hermitian operator since the equality $(\partial_{\alpha}U^{\dagger})U=-U^{\dagger}(\partial_{\alpha}U)$.
Meanwhile,
\begin{equation}
\langle\Delta^{2}\mathcal{H}\rangle_{i}=\langle\psi_{i}|\mathcal{H}^{2}
|\psi_{i}\rangle-\langle\psi_{i}|\mathcal{H}|\psi_{i}\rangle^{2}
\end{equation}
is the variance of $\mathcal{H}$ on the $i$th eigenstate of $\rho_{0}$.
When $\partial_{\alpha}U$ commutes with $U$, $\mathcal{H}$ can
be explained as the generator of the parametrization transformation~\cite{Pang}.
The expression~(\ref{eq:F}) of quantum Fisher information is not
just a formalized representation. The operator $\mathcal{H}$ is only
determined by the parametrization process, that is the dynamics of
the system or the device. For a known dynamical process of a parameter,
i.e., known system's Hamiltonian, $\mathcal{H}$ is a settled operator
and can be obtained in principle. In this representation, the calculation
of quantum Fisher information is separated into two parts: the diagonalization
of initial state and calculation of $\mathcal{H}$. For a general
2-dimensional state, the quantum Fisher information reduces to
\begin{equation}
F_{\mathrm{qubit}}=4\left(2\mathrm{Tr}\rho^2-1\right)\langle\Delta^{2}
\mathcal{H}\rangle_{1(2)}.\label{eq:F_qubit}
\end{equation}
The subscript of the variance can be chosen as 1 or 2 as any Hermitian
operator's variances on two orthonormal states are equivalent in 2-dimentional
Hilbert space. For a pure state, the quantum Fisher information can
be easily obtain from Eq.~(\ref{eq:F_qubit}) with taking the purity $\mathrm{Tr}\rho^2=1$ and the variance on that pure state, i.e.,~\cite{Pang}
\begin{equation}
F_{\mathrm{pure}}=4\langle\Delta^{2}\mathcal{H}\rangle_{\mathrm{in}}.
\end{equation}
Namely, the quantum Fisher information is proportional to the variance of $\mathcal{H}$ on the initial state.
In this scenario, denote the initial state $\rho_{0}=|\psi_{0}\rangle\langle\psi_{0}|$, the quantum
Fisher information can be rewritten into $F_{\mathrm{pure}}=\langle\psi_{0}|L^2_{\mathrm{eff}}|\psi_{0}\rangle$, with the effective SLD operator
\begin{equation}
L_{\mathrm{eff}}=i2\left[\mathcal{H}, |\psi_{0}\rangle\langle\psi_{0}|\right]. \label{eq:L_pure}
\end{equation}

For a well applied form of parametrization transformation $U=\exp(-itH_{\alpha})$~\cite{Pang}, where $\hbar$ has been
set as 1 in Planck unit,  and being aware of the equation
\begin{equation}
\partial_{\alpha}e^{A}=\int_{0}^{1}e^{sA}(\partial_{\alpha}A)e^{(1-s)A}ds,
\end{equation}
$\mathcal{H}$ can then be expressed by
\begin{equation}
\mathcal{H}=-\int_{0}^{t}e^{isH_{\alpha}}\left(\partial_{\alpha}H_{\alpha}\right)e^{-isH_{\alpha}}ds.
\label{eq:H_integral}
\end{equation}
Defining a superoperator $A^{\times}$ as $A^{\times}(\cdot):=[A,\cdot]$,
$\mathcal{H}$ can be written in an expanded form
\begin{equation}
\mathcal{H}=i\sum_{n=0}^{\infty}f_{n}H_{\alpha}^{\times n}\left(\partial_{\alpha}H_{\alpha}\right),\label{eq:H}
\end{equation}
where the coefficient
\begin{equation}
f_{n}=\frac{(it)^{n+1}}{(n+1)!}.
\end{equation}
In many real problems, the recursive commutations in Eq.~(\ref{eq:H}) can either repeat
or terminate~\cite{Jiang14}, indicating an analytical expression
of $\mathcal{H}$. Thus, this representation of quantum Fisher information
would be very useful in these problems. For the simplest case that
$H_{\alpha}=\alpha H$, all terms vanish but the first one, then $\mathcal{H}=-tH$.
When $[H_{\alpha},\partial_{\alpha}H_{\alpha}]=C$, with $C$ a constant
matrix or proportional to $H_{\alpha}$,  only the first and second
terms remain. In this case, $\mathcal{H}$ reduces to $-t(\partial_{\alpha}H_{\alpha}+itC/2)$.
A more interesting case is that $[H_{\alpha},\partial_{\alpha}H_{\alpha}]=c\partial_{\alpha}H_{\alpha}$,
with $c$ a nonzero constant number, then $\mathcal{H}$ can be written
in the form
\begin{equation}
\mathcal{H}=\frac{i}{c}\left[\exp(itc)-1\right]\partial_{\alpha}H_{\alpha}.
\end{equation}

In the following we give an example to exhibit Eq.~(\ref{eq:H}). Consider
the interaction Hamiltonian of a collective spin system in a magnetic field
\begin{equation}
H_{\theta}=B\left(\cos\theta J_{x}+\sin\theta J_{z}\right)=BJ_{\bm{n}_{0}},\label{eq:H_spin}
\end{equation}
where $J_{\bm{n}_{0}}=\bm{n}_{0}\cdot\bm{J}$ with $\bm{n}_{0}=(\cos\theta,0,\sin\theta)^{\mathrm{T}}$
and $\bm{J}=(J_{x},J_{y},J_{z})^{\mathrm{T}}$. $B$ is the amplitude
of the external magnetic field and $\theta$ is the angle between the field and the collective spin. Here $J_{i}=\sum_{k}\sigma_{i}^{(k)}/2$
for $i=x,y,z$ with $\sigma_{i}^{(k)}$ the Pauli matrix for $k$th
spin. Taking $\theta$ as the parameter under estimation, $\mathcal{H}$ can be expressed by
\begin{equation}
\mathcal{H}=2\left|\sin\left(\frac{Bt}{2}\right)\right|J_{\bm{n}_{1}},\label{eq:H_example}
\end{equation}
where $J_{\bm{n}_{1}}=\bm{n}_{1}\cdot\bm{J}$ with the vector
\begin{equation*}
\bm{n}_{1}=\mu\!\left(\cos\!\left(\frac{Bt}{2}\right)\!\sin\!\theta,-\!\sin\!\left(\frac{Bt}{2}\right),
-\!\cos\!\left(\frac{Bt}{2}\right)\!\cos\theta\right)^{\mathrm{T}},
\end{equation*}
where $\mu=\mathrm{sgn}(\sin(Bt/2))$ is the sign function and $\bm{n}_{1}$
is normalized.

The operator $\mathcal{H}$ for Hamiltonian~(\ref{eq:H_spin}) may be also available to be solved using the procedure in Ref.~\cite{Pang}, in the (2j+1)-dimensional eigenspace of $H_{\theta}$ (j is the total spin). In principle, the eigenstates of $H_{\theta}$ can be found by rotating the Dicke state into the same direction of $H_{\theta}$. However, even one can analytically  obtain all the eigenvalues and eigenvectors, it still requires a large amount of calculations to obtain $\mathcal{H}$, especially when the spin numbers are tremendous. Comparably, utilizing Eq.~(\ref{eq:H}), it only takes a few steps of calculation, which can be found in the method. This is a major advantage of the expanded form of $\mathcal{H}$.

\begin{figure}
\includegraphics[width=5cm]{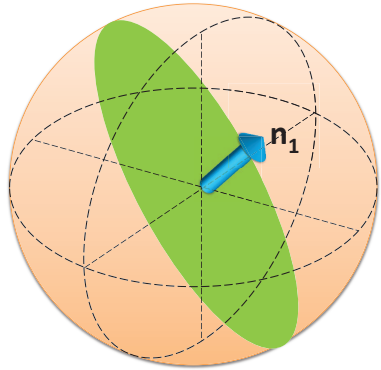}
\caption{\label{fig:sketch} \textbf{Optimal states to access maximum quantum
Fisher information in a spin-half system.} The blue arrow represents
the vector $\bm{n}_{1}$ and all vectors in the green plane are
vertical to $\bm{n}_{1}$. All the states in the joint ring of green plane and Bloch sphere's
surface can access maximum quantum Fisher information.}
\end{figure}

Utilizing Eq.~(\ref{eq:H_example}), one can immediately obtain the
form of $\mathcal{H}$ for a spin-half system
\begin{equation}
\mathcal{H}_{\mathrm{qubit}}=\left|\sin\left(\frac{Bt}{2}\right)\right|\bm{n}_{1}\cdot\bm{\sigma},
\end{equation}
with $\bm{\sigma}=(\sigma_{x},\sigma_{y},\sigma_{z})^{\mathrm{T}}$,
which was also discussed in the Hamiltonian eigenbasis in Ref.~\cite{Pang}.
For any 2-dimensional state, based on Eq.~(\ref{eq:F_qubit}), the
quantum Fisher information can be expressed by
\begin{equation}
F_{\theta}=4\sin^{2}\left(\frac{Bt}{2}\right)|\bm{r}_{\mathrm{in}}|^{2}\left[1-(\bm{n}_{1}\cdot\bm{r}_{\mathrm{e}})^{2}\right],\label{eq:F_spin}
\end{equation}
where $\bm{r}_{\mathrm{in}}=(\langle\sigma_{x}\rangle,\langle\sigma_{y}\rangle,
\langle\sigma_{z}\rangle)^{\mathrm{T}}$
is the Bloch vector of the initial state $\rho_{0}$ and $\bm{r}_{\mathrm{e}}$
is the Bloch vector of any eigenstate of $\rho_{0}$. For pure states,
there is $\bm{r}_{\mathrm{e}}=\bm{r}_{\mathrm{in}}$ and $|\bm{r}_{\mathrm{in}}|=1$.
Since the Bloch vector of a 2-dimensional state satisfies $|\bm{r}_{\mathrm{in}}|\leq1$,
it can be found that the maximum value of Eq.~(\ref{eq:F_spin}) is
\begin{equation}
F_{\theta}^{\mathrm{max}}=4\sin^{2}\left(\frac{Bt}{2}\right),
\end{equation}
which can be saturated when $|\bm{r}_{\mathrm{in}}|=1$ and $\bm{n}_{1}\cdot\bm{r}_{\mathrm{in}}=0$,
namely, the optimal state to access maximum quantum Fisher information
here is a pure state perpendicular to $\bm{n}_{1}$, as shown in Fig.~\ref{fig:sketch}.
In this figure, the yellow sphere represents the Bloch sphere and the
blue arrow represents the vector $\bm{n}_{1}$. It can be found that
all states on the joint ring of the green plane and surface of
Bloch sphere can access the maximum quantum Fisher information, i.e.,
all states on this ring are optimal states. One simple example is
$\bm{r}_{\mathrm{opt}}=\bm{n}_{0}$, and another one is the superposition
state of two eigenstates of $\mathcal{H}$~\cite{Giovannetti06,Pang}.

Alternatively, $B$ could be the parameter that under estimation.
In the spin-half case, with respect to $B$, $\mathcal{H}_{B}=-t\bm{n}_{0}\cdot\bm{\sigma}/2$,
then the quantum Fisher information can be expressed by
\begin{equation}
F_{B}=t^{2}|\bm{r}_{\mathrm{in}}|^{2}\left[1-(\bm{n}_{0}\cdot\bm{r}_{\mathrm{e}})^{2}\right].
\end{equation}
The optimal states to access the maximum value $F_{B}^{\mathrm{max}}=t^{2}$
are the pure states vertical to $\bm{n}_{0}$.

\textbf{Exponential form initial state.} For an exponential form initial state
$\rho_{0}=\exp(G_{0})$, the parametrized state reads
\begin{equation}
\rho_{\alpha}=U\rho_{0}U^{\dagger}=\exp(UG_{0}U^{\dagger}).
\end{equation}
Recently, Jiang~\cite{Jiang14} studied the quantum Fisher information
for exponential states and gave a general form of SLD operator. In
his theory, the SLD operator can be expanded as
\begin{equation}
L=\sum_{n=0}^{\infty}g_{n}G^{\times n}(\partial_{\alpha}G),\label{eq:L}
\end{equation}
where the coefficient
\begin{equation}
g_{n}=\frac{4(2^{n+2}-1)\mathcal{B}_{n+2}}{(n+2)!}
\end{equation}
for even $n$ and $g_{n}$ vanishes for odd $n$. Here $\mathcal{B}_{n+2}$
is the $(n+2)$th Bernoulli number and in our case, $G=UG_{0}U^{\dagger}$.
Through some straightforward calculation, the derivative of $G$ on
$\alpha$ reads
\begin{equation}
\partial_{\alpha}G=-iU\left(G_{0}^{\times}\mathcal{H}\right)U^{\dagger}.
\end{equation}
Based on this equation, the $n$th order term in Eq.~(\ref{eq:L}) is
\begin{equation}
G^{\times n}(\partial_{\alpha}G)=-iU\left[(G_{0}^{\times})^{n+1}\mathcal{H}\right]U^{\dagger}, \label{eq:G_n}
\end{equation}
where $\mathcal{H}$ is given by Eq.~(\ref{eq:H}). Generally, it is known that the quantum Fisher information reads
\begin{equation}
F=\mathrm{Tr}(U\rho_{0}U^{\dagger}L^{2})=\mathrm{Tr}\left(\rho_{0}L_{\mathrm{eff}}^{2}\right),
\end{equation}
where the effective SLD operator $L_{\mathrm{eff}}=U^{\dagger}LU$. The effective SLD operator for pure states is already shown in Eq.~(\ref{eq:L_pure}).
Substituting Eq.~(\ref{eq:G_n}) into Eq.~(\ref{eq:L}), the effective
SLD operator can be expanded as
\begin{equation}
L_{\mathrm{eff}}=-i\sum_{n=0}^{\infty}g_{n}(G_{0}^{\times})^{n+1}\mathcal{H}.\label{eq:L_eff}
\end{equation}

In most mixed states cases, to obtain quantum Fisher information,
the diagonalization of initial state is inevitable, which is the reason
why the usual form of quantum Fisher information is expressed in the
eigenbasis of density matrix. Thus, it is worth to study the expression
of effective SLD operator and quantum Fisher information in the eigenbasis
of $G_{0}$. We denote the $i$th eigenvalue and eigenstate of $G_{0}$
as $a_{i}$ and $|\phi_{i}\rangle$, and in the eigenbasis of $G_{0}$,
the element of $G_{0}^{\times n}\mathcal{H}$ satisfies the recursion
relation
\begin{equation}
[G_{0}^{\times n}\mathcal{H}]_{ij}=(a_{i}-a_{j})[(G_{0}^{\times})^{n-1}\mathcal{H}]_{ij},
\end{equation}
where $[\cdot]_{ij}:=\langle\phi_{i}|\cdot|\phi_{j}\rangle$. Utilizing
this recursive equation, a general formula of $n$th order term can
be obtained,
\begin{equation}
\left[G_{0}^{\times n}\mathcal{H}\right]_{ij}=\left(a_{i}-a_{j}\right)^{n}\mathcal{H}_{ij}.
\end{equation}
Substituting above equation into the expression of $L_{\mathrm{eff}}$
and being aware of the equality
\begin{equation}
\sum_{n=0}^{\infty}g_{n}(a_{i}-a_{j})^{n+1}=2\tanh\left(\frac{a_{i}-a_{j}}{2}\right),
\end{equation}
the element of effective SLD operator in Eq.~(\ref{eq:L_eff}) can be written as
\begin{equation}
[L_{\mathrm{eff}}]_{ij}=-i2\tanh\left(\frac{a_{i}-a_{j}}{2}\right)\mathcal{H}_{ij}.
\end{equation}
Based on the equation $F=\mathrm{Tr}(e^{G_{0}}L_{\mathrm{eff}}^{2})$,
the quantum Fisher information in the eigenbasis of $G_{0}$ can finally
be expressed by
\begin{equation}
F=\sum_{i>j}4\left(e^{a_{i}}+e^{a_{j}}\right)\tanh^{2}\left(\frac{a_{i}-a_{j}}{2}\right)
|\mathcal{H}_{ij}|^{2}.\label{eq:F_sym}
\end{equation}
This is one of the main results in this paper.
In some real problems, the eigenspace of $G_{0}$ could be find easily.
For instance, the eigenspace of a bosonic thermal state is the Fock
space. Thus, as long as the formula of $\mathcal{H}$ in Fock space
is established, the quantum Fisher information can be obtained from
Eq.~(\ref{eq:F_sym}).

Now we exhibit Eq.~(\ref{eq:F_sym}) with a spin-half
thermal state. The initial state is taken as
\begin{eqnarray}
\rho_{0}&=&\frac{1}{Z}\exp\left(-\beta\sigma_{z}\right), \notag \\
&=& \exp\left(-\beta \sigma_z-\ln Z\right),
\end{eqnarray}
where $\beta=1/(k_{\mathrm{b}}T)$ with $k_{\mathrm{b}}$ the Boltzmann constant and $T$ the temperature.
In Planck unit, $k_{\mathrm{b}}=1$. The partition function reads $Z=\mathrm{Tr}[\exp(-\beta\sigma_{z})]=2\cosh\beta$.
In this case, $G_{0}=-\beta \sigma_z-\ln z$.
Denoting the eigenstates of $\sigma_{z}$ as $|0\rangle$ and $|1\rangle$,
i.e., $\sigma_{z}=|0\rangle\langle0|-|1\rangle\langle1|$, the eigenvalues of
$G_{0}$ read $a_{1}=-\beta\sigma_{z}-\ln z$ and $a_{2}=\beta\sigma_{z}-\ln z$.
The parametrization process is still taken as $H_{\theta}=B\bm{n}_{0}\cdot\bm{\sigma}/2$ with $\theta$ the parameter under estimation, indicating that $\mathcal{H}=\left|\sin\left(\frac{Bt}{2}\right)\right|\bm{n}_{1}\cdot\bm{\sigma},$
then the squared norm of the off-diagonal element of $\mathcal{H}$
in the eigenbasis of $\sigma_{z}$ reads
\begin{equation}
|\mathcal{H}_{01}|^{2}=\sin^{2}\left(\!\frac{Bt}{2}\!\right)\!\!
\left[1-\cos^{2}\theta\cos^2\left(\frac{Bt}{2} \right)\right].
\end{equation}
Immediately, the quantum Fisher information can be obtained from Eq.~(\ref{eq:F_sym}) as
\begin{equation}
F_{\mathrm{T}}=4\tanh^{2}\beta\sin^{2}\!\!\left(\!\frac{Bt}{2}\!\right)
\!\!\left[1-\cos^{2}\theta\cos^2\left(\frac{Bt}{2} \right) \right]. \label{eq:F_thermal}
\end{equation}
The maximum value of above expression is obtained at $Bt=(4k+1)\pi$
for $k=0,1,...$ and
\begin{equation}
F_{\mathrm{T}}^{\mathrm{max}}=4\tanh^{2}(\beta).
\end{equation}

\begin{figure}
\includegraphics[width=7.5cm]{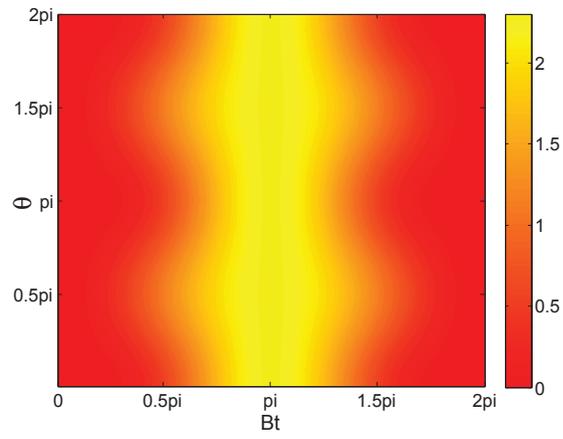}
\protect\caption{\label{fig:thermal}\textbf{Quantum Fisher information as a function
of} $Bt$ \textbf{and} $\theta$. The initial state is a spin-half
thermal state and the temperature is set as $T=1$ here. }
\end{figure}

From this equation, one can see that the value of maximum quantum
Fisher information is only affected by the temperature. With the increase
of temperature, the maximum value reduces. In the other hand, quantum Fisher
information in Eq.~(\ref{eq:F_thermal}) is related to $Bt$ and $\theta$.
Fig.~\ref{fig:thermal} shows the quantum Fisher information as a function of $Bt$ and $\theta$.
The values of $Bt$ and $\theta$ are both within $[0,2\pi]$ in the
plot. The temperature is set as $T=1$ here. From this figure, it
can be found that the maximum quantum Fisher information is robust
for $\theta$ since it is always obtained at $Bt=\pi$ for any value
of $\theta$. Furthermore, this optimal condition of $Bt$ is independent
of temperature. With respect to $Bt$, there is a large regime near
$Bt=\pi$ in which the quantum Fisher information's value can surpass
$2$, indicating that the quantum Fisher information can be still
very robust and near its maximum value even when $Bt$ is hard to
set exactly at $\pi$.

\textbf{Multiparameter processes.} For a multiparameter system, the
element of quantum Fisher information matrix in Ref.~\cite{PhysicaAliu} can also be written with
$\mathcal{H}$ operator,
\begin{eqnarray}
\mathcal{F}_{\alpha\beta} \!\!&=&\!\! \sum_{i=1}^{M}4p_{i}\mathrm{cov}_{i}\left(\mathcal{H}_{\alpha},\mathcal{H}_{\beta}\right)\nonumber \\
&  & -\sum_{i\neq j}\frac{8p_{i}p_{j}}{p_{i}+p_{j}}\!\mathrm{Re}\!\left(\langle\psi_{i}|\mathcal{H}_{\alpha}
|\psi_{j}\rangle\langle\psi_{j}|\mathcal{H}_{\beta}|\psi_{i}\rangle\right)\!, \label{eq:F_alpha_beta}
\end{eqnarray}
where $U$ is dependent on a series of parameters $\alpha$, $\beta$
and so on, and
\begin{equation}
\mathcal{H}_{m}=i(\partial_{m}U^{\dagger})U,
\end{equation}
with the index $m=\alpha,\beta,...$. The covariance matrix on the
$i$th eigenstate of initial state is defined as
\begin{equation*}
\mathrm{cov}_{i}(\mathcal{H}_{\alpha},\mathcal{H}_{\beta})\!:=\!
\frac{1}{2}\langle\psi_{i}|\!\{\!\mathcal{H}_{\alpha},\mathcal{H}_{\beta}\!\}
\!|\psi_{i}\rangle-\langle\psi_{i}|\!\mathcal{H}_{\alpha}\!|\psi_{i}\rangle\langle\psi_{i}|
\!\mathcal{H}_{\beta}\!|\psi_{i}\rangle,
\end{equation*}
with $\{\cdot,\cdot\}$ the anti-commutation. For a single qubit system, Eq.~(\ref{eq:F_alpha_beta}) reduces to
\begin{equation}
\mathcal{F}_{\mathrm{qubit},\alpha\beta}= 4\left(2\mathrm{Tr}\rho^2-1\right)\mathrm{cov}_{1(2)}\left(\mathcal{H}_{\alpha},\mathcal{H}_{\beta}\right).
\label{eq:F_multi}
\end{equation}
Similarly with the single-parameter scenario, the subscript in Eq.~(\ref{eq:F_multi})
can be chosen as 1 or 2 since the covariance for two Hermitian
operators are the same on two orthonormal states in 2-dimensional
Hilbert space. From this equation, the element of quantum Fisher information
matrix for pure states can be immediately obtained as
\begin{equation}
\mathcal{F}_{\mathrm{pure},\alpha\beta}=4 \mathrm{cov}_{\mathrm{in}}\left(\mathcal{H}_{\alpha},\mathcal{H}_{\beta}\right),
\end{equation}
namely, for pure states, the element of quantum Fisher information
matrix is actually the covariance between two $\mathcal{H}$ operators on the initial state. When the total Hamiltonian can be written as $\sum_{i}\alpha_{i}H_{i}$ and $[H_{i},H_{j}]=0$ for any $i$, $j$, above equation can reduce to the covariance between $H_{i}$ and $H_{j}$~\cite{Liu_H}. For the diagonal elements, they are exactly the quantum Fisher information for the corresponding parameters.

For multiparamter estimations, the Cram\'{e}r-Rao bound cannot always be achieved. In the scenario of pure states, the condition of this bound to be tight is
$\mathrm{Im}\langle\psi_{\mathrm{out}}|L_{\alpha}L_{\beta}
|\psi_{\mathrm{out}}\rangle=0$, $\forall \alpha,\beta$~\cite{Matsumoto,Fujiwara}.
Here $|\psi_{\mathrm{out}}\rangle$ is dependent on the parameter under estimation. In the unitary parametrization, $|\psi_{\mathrm{out}}\rangle=U|\psi_{0}\rangle$ and this condition can be rewritten into $\mathrm{Im}\langle\psi_{0}|L^{\alpha}_{\mathrm{eff}}L^{\beta}_{\mathrm{eff}}
|\psi_{\mathrm{0}}\rangle=0$, $\forall \alpha,\beta$. Here $L^{\alpha(\beta)}_{\mathrm{eff}}=U^{\dagger}L_{\alpha(\beta)}U$ is the effective SLD operator for parameter $\alpha(\beta)$. Utilizing Eq.~(\ref{eq:L_pure}), this condition can be expressed in the form of $\mathcal{H}$ operator,
\begin{equation}
\langle\psi_{0}|\left[\mathcal{H}_{\alpha}, \mathcal{H}_{\beta} \right]|\psi_{0}\rangle=0, \quad\forall \alpha,\beta. \label{eq:CR_condition}
\end{equation}
In other word, $\langle\psi_{0}|\mathcal{H}_{\alpha} \mathcal{H}_{\beta}|\psi_{0}\rangle$ needs to be a real number for any $\alpha$ and $\beta$. When $\mathcal{H}_{\alpha}$ commutes with $\mathcal{H}_{\beta}$ for any $\alpha$ and $\beta$, above condition can always be satisfied for any initial state.

Generally, for the unitary parametrization process, the element of quantum Fisher information matrix can be expressed by $F=\mathrm{Tr}(\rho \left\{L_{\alpha},L_{\beta}\right\})=\mathrm{Tr}(\rho_{0}\{L^{\alpha}_{\mathrm{eff}},
L^{\beta}_{\mathrm{eff}}\})$. From the definition equation of SLD, one can see that $L^{\alpha}_{\mathrm{eff}}$ satisfies the equation $\partial_{\theta}\rho=U\{\rho_{0},L_{\mathrm{eff}}\}U^{\dagger}/2$. The quantum Fisher information matrix has more than one definitions. One alternative candidate is using the so-called Right Logarithmic Derivative (RLD)~\cite{Holevo,Belavkin,Kim}, which is defined as $\partial_{\alpha}\rho=\rho R_{\alpha}$, with $R_{\alpha}$ the RLD. The element of RLD quantum Fisher information matrix can be written as
\begin{equation}
\mathcal{J}_{\alpha\beta}=\mathrm{Tr}\left(\rho R_{\alpha}R^{\dagger}_{\beta}\right)=\mathrm{Tr}\left(\rho_{0} R^{\alpha}_{\mathrm{eff}} R^{\beta \dagger}_{\mathrm{eff}}  \right),
\end{equation}
where the effective RLD reads $R^{\alpha(\beta)}_{\mathrm{eff}}=U^{\dagger}R_{\alpha(\beta)}U$.
For a unitary parametrization process, assuming the initial state has nonzero determinant, $R^{\alpha}_{\mathrm{eff}}$ can be expressed by $\mathcal{H}_{\alpha}$ and the initial state $\rho_{0}$, i.e.,
\begin{equation}
R^{\alpha}_{\mathrm{eff}}=i\left(\rho^{-1}_{0}\mathcal{H}_{\alpha}\rho_{0}
-\mathcal{H}_{\alpha}\right).
\end{equation}
With this equation, the element of RLD quantum Fisher information matrix can be expressed by
\begin{equation}
\mathcal{J}_{\alpha\beta}=\mathrm{Tr}(\mathcal{H}_{\alpha}\rho^{2}_{0}\mathcal{H}_{\beta}
\rho^{-1}_{0}-2\mathcal{H}_{\beta}\mathcal{H}_{\alpha}\rho_{0}
+\mathcal{H}_{\alpha}\mathcal{H}_{\beta}\rho_{0}).
\end{equation}
When the parametrization process is displacement, this equation can reduces to the corresponding form in Ref.~\cite{Kim}. For pure states, the element reads
$\mathcal{J}_{\mathrm{pure},\alpha\beta}=\mathrm{Tr}\left[\left(\partial_{\alpha}\rho \right) \left(\partial_{\beta}\rho\right)\right]
=\mathcal{F}_{\mathrm{pure},\alpha\beta}/2$.
Recently, Genoni \emph{et al.}~\cite{Kim} proposed a most informative Cram\'{e}r-Rao bound for the total variance of all
parameters under estimation. From the relation between $\mathcal{J}_{\mathrm{pure},\alpha\beta}$ and $\mathcal{F}
_{\mathrm{pure},\alpha\beta}$, one can see that $\mathrm{Tr}\mathcal{F}^{-1}_{\mathrm{pure}}$ is always larger than $
\mathrm{Tr}\mathcal{J}^{-1}_{\mathrm{pure}}$, namely, the SLD Cram\'{e}r-Rao bound is always more informative than the
RLD counterpart in this scenario.

We still consider the spin-half system with the Hamiltonian $H=B\bm{n}_{0}\cdot\bm{\sigma}/2$.
Take both $B$ and $\theta$ as the parameters under estimations.
First, based on aforementioned calculation, the $\mathcal{H}$ operator
for $B$ and $\theta$ read
\begin{eqnarray}
\mathcal{H}_{B} & = & -\frac{t}{2}\bm{n}_{0}\cdot\bm{\sigma},\\
\mathcal{H}_{\theta} & = & \left|\sin\left(\frac{Bt}{2}\right)\right|\bm{n}_{1}\cdot\bm{\sigma}.
\end{eqnarray}
Based on the property of Pauli matrices  $\{\bm{n}_{0}\cdot\bm{\sigma},\bm{n}_{1}\cdot\bm{\sigma}\}=2\bm{n}_{0}\cdot\bm{n}_{1}$,
the anti-commutation in the covariance reads
\begin{equation}
\left\{ \mathcal{H}_{B},\mathcal{H}_{\theta}\right\} =-t\left|\sin\left(\frac{Bt}{2}\right)\right|\bm{n}_{0}\cdot\bm{n}_{1}.
\end{equation}
For a pure initial state, the off-diagonal element of the quantum
Fisher information matrix is expressed by
\begin{equation}
\mathcal{F}_{B\theta}=2t\left|\sin\left(\frac{Bt}{2}\right)\right|\left(\bm{n}_{0}
\cdot\bm{r}_{\mathrm{in}}\right)\left(\bm{n}_{1}\cdot\bm{r}_{\mathrm{in}}\right), \label{eq:off-diagonal}
\end{equation}
where $\bm{r}_{\mathrm{in}}$ is the Bloch vector of the initial pure
state and the equality $\bm{n}_{0}\cdot\bm{n}_{1}=0$ has been used. When the initial pure state is vertical to $\bm{n}_{0}$ or $\bm{n}_{1}$, this off-diagonal element vanishes. Compared with the optimal condition for maximum quantum Fisher information for $B$ and $\theta$ individually, the Bloch vector $\bm{n}_{2}=\bm{n}_{0}\times\bm{n}_{1}$ can optimize both the diagonal elements of quantum Fisher information matrix and vanish the off-diagonal elements. However, all above is only necessary conditions for the achievement of Cram\'{e}r-Rao bound. To find out if the bound can be really achieved, the condition~(\ref{eq:CR_condition}) needs to be checked. In this case,
\begin{equation}
\left[\mathcal{H}_{B}, \mathcal{H}_{\theta}\right]=-it\left|\sin\left(\frac{Bt}{2}\right)\right|
\bm{n}_{2}\cdot\bm{\sigma}.
\end{equation}
With this equation, condition~(\ref{eq:CR_condition}) reduces to $\bm{n}_{2}\cdot\bm{r}_{\mathrm{in}}=0$, i.e., to make the Cram\'{e}r-Rao bound achievable, the Bloch vector of the initial state needs to in the plane of $\bm{n}_{0}$ and $\bm{n}_{1}$. Unfortunately, $\bm{n}_{2}$ is not in this plane. Thus, $B$ and $\theta$ cannot be optimally joint measured simultaneously.

In the plane constructed by $\bm{n}_{0}$ and $\bm{n}_{1}$, any Bloch vector of pure state can be written as
$\bm{r}_{\mathrm{in}}=\bm{n}_{0}\cos\phi+\bm{n}_{1}\sin\phi$, then we have
$\mathcal{F}_{BB}=t^2\sin^2\phi$, $\mathcal{F}_{\theta\theta}=4\sin^2(Bt/2)\cos^2\phi$, and $\mathcal{F}_{B\theta}=2t|\sin(Bt/2)|\cos\phi\sin\phi$. From these expressions, one can see that the determinant of quantum Fisher information matrix is zero, i.e., $\det\mathcal{F}=0$. This fact indicates that, utilizing any pure state in this plane, the variances of $B$ and $\theta$ cannot be estimated simultaneously through the Cram\'{e}r-Rao theory.

\vspace{8pt}
\noindent\textbf{Discussion}\\
We have discussed the quantum Fisher information with unitary parametrization
utilizing an alternative representation. The total information of
the parametrization process is involved in a $\mathcal{H}$
operator in this representation. This operator is totally determined by the parameter and parametrization transformation $U$. As long as the parameter and transformation are taken, $\mathcal{H}$ is a settled operator and independent of the initial state. More interestingly, $\mathcal{H}$ can be expressed in an expanded form. For the Hamiltonians owning recursive commutations with their partial derivative on the parameter under estimation, this expanded form shows a huge advantage. Utilizing this representation,
we give a general analytical expression of quantum Fisher information
for an exponential form initial state. Moreover, we have also studied the $\mathcal{H}$ representation in multiparameter processes. The condition of Cram\'{e}r-Rao bound to be achievable for pure states are also presented in the form of $\mathcal{H}$ operator. In addition, we give the $\mathcal{H}$ representation of Right Logarithmic Derivative and the corresponding quantum Fisher information matrix.

As a demonstration, we apply this representation in a collective spin
system and show the expression of $\mathcal{H}$. Furthermore,
we provide an analytical expression of quantum Fisher information
in a spin-half system. If we consider this system as a multiparameter
system, the corresponding quantum Fisher information matrix can also
be straightforwardly obtained by this representation. From these
expressions, one can find the optimal states to access the maximum
quantum Fisher information. For the parameter $B$, the optimal state
is a pure state vertical to $\bm{n}_{0}$, and for the parameter $\theta$,
the optimal one is also a pure state, but vertical to $\bm{n}_{1}$.
By analyzing the off-diagonal element of quantum Fisher information
matrix, the states to optimize the diagonal elements and make the off-diagonal elements vanish are found. However, these states fail to satisfy the condition of achievement. Thus, $B$ and $\theta$ cannot be optimally jointed measured.

\vspace{8pt}
\noindent\textbf{Methods}\\
\textbf{Collective spin system in a magnetic field.} For the Hamiltonian~(\ref{eq:H_spin}),
its derivative on parameter $\theta$ is $\partial_{\theta}H_{\theta}=\bm{n}_{0}^{\prime}\cdot\bm{J}=J_{\bm{n}_{0}^{\prime}}=-iH_{\theta}^{\times}J_{y}$
with the vector $\bm{n}_{0}^{\prime}=d\bm{n}_{0}/d\theta=(-\sin\theta,0,\cos\theta)^{\mathrm{T}}$.
Based on Eq.~(\ref{eq:H}), $\mathcal{H}$ can be written as
\begin{equation}
\mathcal{H}=\left[\exp\left(itH_{\theta}^{\times}\right)-1\right]J_{y}.
\end{equation}
It is worth to notice that $H_{\theta}^{\times}=BJ_{\bm{n}_{0}}^{\times}$,
then $\mathcal{H}$ is
\begin{equation}
\mathcal{H}=\left [\exp\left(iBtJ_{\bm{n}_{0}}^{\times}\right)-1\right ]J_{y}.
\end{equation}
Being aware of the commutation relations
\begin{eqnarray}
[J_{\bm{n}_{0}},J_{y}] &=& iJ_{\bm{n}_{0}^{\prime}},\\
{}[J_{\bm{n}_{0}},J_{\bm{n}_{0}^{\prime}}] &=& -iJ_{y},
\end{eqnarray}
one can straightforwardly obtain the $n$th order term as below
\begin{equation}
J_{\bm{n}_{0}}^{\times n}J_{y}=\begin{cases}
J_{y}, & \mathrm{for}\,\mathrm{even}\, n;\\
iJ_{\bm{n}_{0}^{\prime}}. & \mathrm{for}\,\mathrm{odd}\, n.
\end{cases}
\end{equation}
With this equation, $\mathcal{H}$ can be expressed by
\begin{equation}
\mathcal{H}=[\cos\left(Bt\right)-1]J_{y}-\sin\left(Bt\right)J_{\bm{n}_{0}^{\prime}},
\end{equation}
equivalently, it can be written in a inner product form: $\mathcal{H}=\bm{r}\cdot\bm{J}$,
where the elements of $\bm{r}$ read $r_{x}=\sin(Bt)\sin\theta$,
$r_{y}=\cos(Bt)-1$ and $r_{z}=-\sin(Bt)\cos\theta$. After the normalization
process, $\mathcal{H}$ is rewritten into the form of Eq.~(\ref{eq:H_example}).

For a spin-half system, the quantum Fisher information can be expressed by
\begin{equation}
F=4\sin^{2}\left(\frac{Bt}{2}\right)|\bm{r}_{\mathrm{in}}|^{2}\left[1-(\bm{n}_{1}
\cdot\langle\bm{\sigma}\rangle_{1(2)})^{2}\right],
\end{equation}
where $\bm{r}_{\mathrm{in}}$ is the Bloch vector of $\rho_{0}$ and
can be obtained through the equation
\begin{equation}
\rho_{0}=\frac{1}{2}\openone+\frac{1}{2}\sum_{i=x,y,z}r_{\mathrm{in},i}\sigma_{i},
\end{equation}
with $\openone$ the identity matrix. $\langle\bm{\sigma}\rangle_{i}=(\langle\sigma_{x}\rangle_{i},\langle\sigma_{y}
\rangle_{i},\langle\sigma_{z}\rangle_{i})^{\mathrm{T}}$
is the vector of expected values on the $i$th ($i=1,2$) eigenstate
of $\rho_{0}$. It can also be treated as the Bloch vector of the eigenstates.
In previous sections, we denote $\bm{r}_{\mathrm{e}}:=\langle\bm{\sigma}\rangle_{i}$.

\vspace{8pt}
\noindent\textbf{Acknowledgments}\\
The authors thank Dr. X.-M. Lu for helpful discussion. This work was supported by the NFRPC through Grant No. 2012CB921602 and the NSFC through Grants No. 11475146.

\vspace{8pt}
\noindent\textbf{Author contributions}\\
X.W. and J.L. contributed the idea. J.L. performed the calculations and prepared the figures.
X.J. checked the calculations. J.L. wrote the main manuscript and X.W. made an improvement.
All authors contributed to discussion and reviewed the manuscript.

\vspace{8pt}
\noindent\textbf{Additional information}\\
Competing financial interests: The authors declare no competing financial interests.

\end{document}